# An Asymmetric Spoof-Fluid-Spoof Acoustic Waveguide and its Application as a CO$_2$ Sensor

Nathan Perchikov[1], Đorđe Vujić[1], Branimir Bajac[1], Andrea Alù[2], Vesna Bengin[1], and Nikolina Janković[1]

[1] BioSense Institute, University of Novi Sad, Novi Sad, Serbia
[2] Advanced Science Research Center, City University of New York, New York, USA

*To whom correspondence should be addressed: nathan.perchikov@biosense.rs, nikolina@biosense.rs

**Abstract**

We study pressure acoustic propagation in asymmetric spoof-fluid-spoof acoustic waveguides and its potential application in acoustic gas sensors. First, a stable and efficient analytical method is established for fast calculation of the dispersion curves based on spectral expansion and enforcement of continuity between segments at suitable collocation points. The analysis is validated by a commercial finite element software. The geometric design of the waveguide is then optimized for the emergence of a nearly-flat dispersion curve associated with vertical geometric asymmetry. The waveguide is fabricated using 3D printing technology and the measurement results corroborate the numerical simulations. Based on the nearly-flat dispersion curve supported by this waveguide, a CO$_2$ sensor is proposed allowing to relate the phase difference measured between two points in the waveguide to the composition of the gas in the waveguide. The proposed sensor is experimentally validated in a controlled environment and the measurement results match the computational predictions well. The sensor is robust with respect to noise and signal-recording duration due to fast phase measurements and shows high sensitivity to gas concentration due to reliance on the second, nearly-flat, dispersion curve. In addition, the sensor is label-free and low-cost, while exhibiting rapid response, low-maintenance requirements and potential for measurements in a wide range of CO$_2$ concentrations without saturation issues.

## 1. Introduction

Periodic structures have been studied for decades, however only with the introduction of the concepts of phononic and photonic crystals and later metamaterials, have these structures unlocked their full potential allowing for almost arbitrary wave manipulation and the engineering of devices with novel functionalities [1-8]. Subwavelength unit cells and the presence of local resonances are the key features of metamaterials that provide properties radically different from those of conventional media. Owing to these features, spoof surface plasmon polaritons (SSPP), which emulate natural surface plasmon polaritons albeit operating in other frequency ranges than the optical one, have been successfully developed [9-16].

Based on the same idea, acoustic SSPP have been realized at the interface of a fluid and a corrugated surface, exhibiting similar properties in terms of wave localization [17-21]. The phenomenon of acoustic SSPP has found a number of applications, including collimation of sound [22-24], bending of sound [25], extraordinary transmission [26], focusing [27–29], imaging [29], multiband propagation [30], rainbow trapping [31], filtering and sensing [32].

Recently, we studied an aSSPP waveguide that is a counterpart of a metal-insulator-metal plasmonic waveguide, i.e., a spoof-fluid-spoof (SFS) acoustic waveguide [33]. Although the structure has a potential for various applications, it should be noted that the SFS waveguide represented a symmetric case, in which the two surfaces had the same geometry. The performed experiments validated the dispersion characteristics of the first two modes as calculated, albeit without demonstrating any particular applications.

One of the applications for which acoustic structures show a great potential is gas sensing, which is critical for monitoring and detecting the concentration of hazardous gases, air quality, environmental protection, or general analysis across various industries, including safety, aerospace, life-science, and the medical, and agricultural industries [34-39]. The main advantages of acoustic structures when compared to other types of gas sensors are robustness, low maintenance, rapid response, and, generally, low cost.

The first acoustic gas sensors were based on speed-of-sound measurements and the associated sensing platforms were typically bulky, as they included acoustic pipes and large actuators and transducers [40-44]. Significant size reduction was achieved with sensors that employed piezoelectric materials and propagation of various surface (SAW) and bulk acoustic waves (BAW) such as the Rayleigh, Love, Lamb, bulk-longitudinal, thickness-shear, and thickness-extension modes [45-51]. While being very attractive due to their size and performance, such sensors usually require advanced fabrication technologies.

More recent types of acoustic sensors usually rely on periodic structures, the typical examples being phononic crystals [52-60] and spoof-plasmon-based sensors [61]. It should be said, however, that the majority of the proposed sensors were only studied analytically and numerically, never providing experimental verification, which is crucial for practical applications. Moreover, in most of the studies the sensor-response is based on damped-resonance amplitude-change, rather than on phase-measurement. In addition, only the fundamental mode of the structure is utilized, which exhibits linear behavior in a significant portion of the wavenumber range. Both choices limit the sensitivity potential. Another issue is that pressure-amplitude measurements are more susceptible to noise, when compared to phase measurements, and thus the latter can provide more robustness and reliability.

An advantageous alternative is the utilization of nearly-symmetric structures, in which the symmetry is associated with the possibility to excite perfectly-localized modes. Specifically, the higher-mode dispersion bands of nearly-symmetric periodic structures can be nearly-flat, which can provide significantly larger sensitivity for a sensor in comparison to sensors based on the fundamental mode dispersion. In general, higher modes and the related flat bands have been widely investigated both in purely periodic structures and in periodic structures with defects [62-71], however, the utilization of these phenomena remained unexploited in sensing structures.

In this paper, we first extend the concept of symmetric SFS waveguides to a general asymmetric case, in which the two bounding surfaces have different geometrical parameters. A thorough spectral point-collocation-based dispersion analysis along with finite-element analysis validation are presented. We show that one of the main advantages of such a structure, with the additional degrees of freedom, is the possibility to manipulate higher modes in a more flexible way. More concretely, the first higher mode can be tailored to have a nearly-flat dispersion diagram, which consequently produces a significant phase change even for minute changes in the surrounding fluid characteristics. Afterwards, using nearly-flat mode as the underlying idea, we propose a very sensitive acoustic gas sensor for the detection of $CO_2$. After the numerical analysis and the parametric optimization of the design, the sensor is fabricated using low-cost 3D printing technology and then experimentally tested in a controlled environment. Besides being cheap and easily fabricated, the proposed sensor has the advantage of rapid response without the need for recovery or maintenance, higher robustness and sensitivity when compared to other acoustic sensors as well as successful operation in a large gas-concentrations range.

To the best of our knowledge, this is the first sensor based on the concept of relating fluid characteristics to acoustic phase-difference measurements using for waves excited at frequencies associated with the higher, nearly-flat dispersion mode. The proposed sensor can be further extended to perform with other types of analytes.

The structure of this paper is as follows: after the Introduction, in Section 2 and the associated Appendix A, we present the theoretical background and the employed computational framework, which includes point-collocation-based spectral analysis, as well as finite-element simulations. Section 3 exhibits experimental validation of the numerical results. Section 4 details how the proposed structure can be utilized for gas-sensing applications. Section 5 gives a thorough overview of the fabrication and measurements results for the proposed $CO_2$ sensor. Further comparative analysis of the quality and novelty of the sensor is given in Section 6. Section 7 concludes.

## 2. Theoretical Background

The layout of the proposed asymmetric spoof-fluid-spoof (ASFS) waveguide, along with the corresponding geometrical parameters, are shown in Figure 1. We consider lossless two-dimensional geometry with perfect rigid surfaces, the structure being invariant in the $z$ direction and bounded in the $y$ direction. The ASFS waveguide is vertically asymmetrical, consisting of two surfaces with corrugations that in a general case have different geometrical parameters but the same periodicity $d$ and are separated by a gap $g$. The fluid that occupies the structure is characterized by its density $\rho$ and speed of sound $C$.

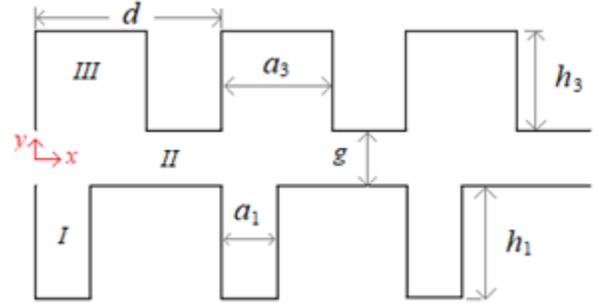

Figure 1: A 2D sketch of the asymmetric spoof-fluid-spoof waveguide.

There are two main computational approaches to obtain the dispersion diagram of the proposed structure. The first approach is finite element discretization with iterative solution of the associated linear eigenvalue problem combined with the mode-matching technique, where continuity conditions are enforced between the different rectangular segments in the waveguide in the integral sense, using projection on the basis-function space [72]. Another approach is the point-collocation method, following which continuity is enforced on a given set of discrete points [73].

The finite-element approach basically approximates the unknown functions using a basis of piece-wise linear functions with so-called local support [74]. The advantage of the finite-element approach is that increase in resolution does not automatically translate to increase in the condition number of the representative matrix of the eigenvalue problem associated with the dispersion analysis. On the other hand, special care has to be taken to avoid spurious (non-wave) solutions. In contrast, if one expands the unknown fields individually in each rectangular segment using a basis of wave functions, the physicality of the solution is guaranteed. However, due to the fact that waves are described by complex exponents, the approach suffers from certain numerical stiffness, since increase in resolution does directly translate to increase in the condition number of the matrix representing the continuity of fields between the different segments. For instance, in [75] a mode-matching strategy was employed, entailed in projecting every basis function of the wave-equation kernel by use of a basis of similar wave functions, albeit with a different period, the ratio of the periods being dependent on the respective sizes of the rectangular segments comprising the waveguide periodicity-domain. Such an approach leads to very large representative continuity-enforcing matrices. Consequently, for highly asymmetric structures, the condition number of the representative matrix may become so high as to make the problem intractable on a regular processor.

In contrast, the present work proposes to employ the point-collocation strategy, as an alternative for the mode-matching technique, while remaining in the framework of the wave basis-functions approach. One disadvantage of the point-collocation approach is the high sensitivity of the solution to the choice of the collocation points. However, if a robust choice of points is suggested, the approach becomes advantageous. A detailed derivation of the dispersion equation using the point-collocation approach is given in Appendix A together with a thorough explanation related to the stability of the approach and the choice of the collocation points.

To validate the spectral analysis, dispersion diagrams for the proposed structure have also been obtained using the commercial finite-element program COMSOL Multiphysics using the two-dimensional pressure-acoustics frequency-domain eigenfrequency solver. Figure 2(a) shows a dispersion diagram for the first five modes obtained using the spectral-collocation and the finite-element methods for the ASFS structure with $a_1$ = 10.3 mm, $a_3$ = 10.3 mm, $d$ = 34.3 mm, $g$ = 10.3 mm, $h_1$=18.2 mm, $h_3$ =20.6 mm, confirming excellent agreement between the two methods. Figure 2(b) shows the acoustic (Fourier-space) pressure for the second-mode middle wavenumber in a unit cell, and one can note typical even and odd field distributions, respectively. Figures 3 and 4 show dispersion diagrams for the first two modes for the ASFS structure with $a_1$ = 10.3 mm, $a_3$ = 10.3 mm, $d$ = 34.3 mm, $g$ = 10.3 mm, where the parameters $h_1$ and $h_3$ are varied. The fluid in the structure has the following parameters: $\rho$ = 1.1802$kg/m^3$, $C$ = 346.8705$m/s$.

Excellent agreement between the analytical and numerical results can be observed in Figure 2, confirming the validity of the chosen analytical approach. The fundamental mode exhibits the expected behavior, i.e., it is predominantly linear, where the slope of the flat part can be modulated by the change of the geometrical parameters.

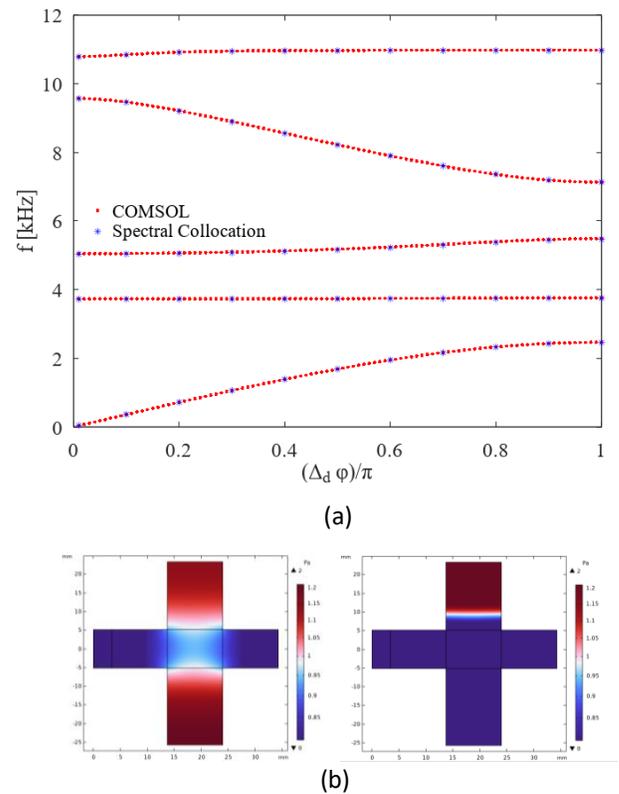

Figure 2: (a) Dispersion diagram of the first five modes for $a_1$ = $a_3$ = $g$ = 10.3 mm, $d$ = 34.3 mm, $h_1$=18.2 mm, $h_3$ =20.6 mm, obtained using the spectral-collocation and the finite-element methods; (b) acoustic (Fourier-space) pressure (in logarithmic color scale) for the first- and second-mode middle wavenumbers (left and right, respectively) in a unit cell, exhibiting even and odd field distributions.

However, of particular interest is the second mode, since the asymmetry in the structure enables flexible manipulation of its slope. In other words, the configuration can be used to tailor the second mode to be nearly-flat, i.e., to have slow-wave nature yet enabling propagation and such a property can be of a great advantage in the design of a sensor because it can provide fast sensing and high sensitivity.

In other words, in the scenario of quasi-flatness and a sensing principle based on a phase change, a minute change of a fluid parameter would not change the *shape* of the dispersion diagram, but it would slightly push it towards higher frequencies. Measured at a specific frequency, the signals before and after the minute change in the fluid would have significantly different wavenumbers. This is the basic idea of the sensor, which will be elaborated on in the following. While the same principle can be applied to the fundamental mode, it should be noted that the linearity or the high slope of the fundamental mode for most wavenumbers prevents it from providing significantly different wavenumbers for minute changes in the fluid at a given excitation frequency, as supported by the nearly-flat mode.

We observe from comparing Figures 3(a) and 4(a) that vertical perturbation of the symmetry is a higher-order effect than the horizontal perturbation. Both effects are small, and small perturbation does not affect the first mode much. But a small change in $h_1/h_3$ gives a large change already in the *second* mode dispersion curve, whereas a small change in $a_3/a_1$ gives a large change only in the *third* mode of the dispersion curve.

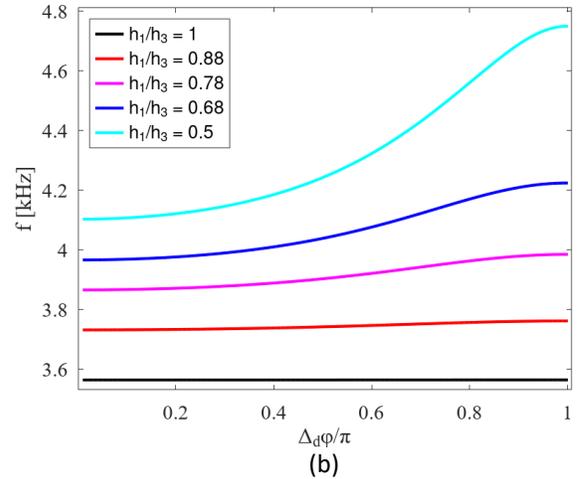

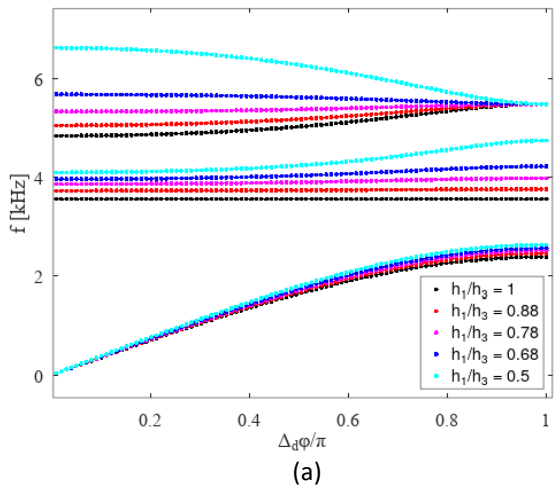

Figure 3: Dispersion diagrams of the first three modes (a) and a zoom-in on the second mode (b) for $a_1 = a_3 = g = 10.3$ mm, $d = 34.3$ mm, $h_3 = 20.6$, and various $h_1/h_3 <1$ values.

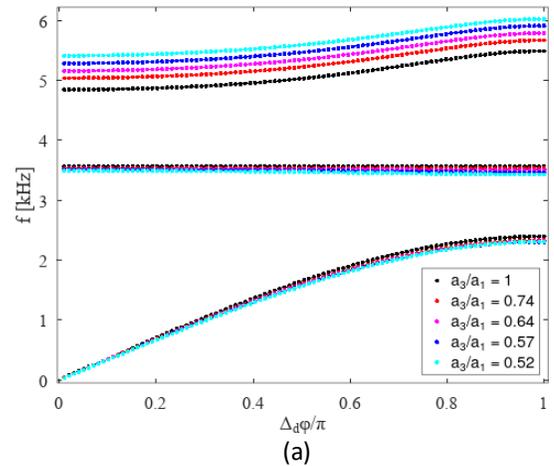

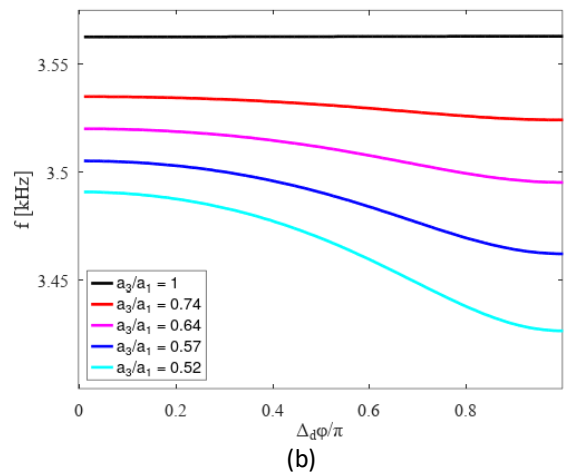

Figure 4: Dispersion diagrams of the first three modes (a) and a zoom-in on the second mode (b) for $h_1 = h_3 = 20.6$mm, $g = 10.3$ mm, $d = 34.3$ mm, $a_3 = 10.3$, and various $a_3/a_1 <1$ values.

This suggests that using a vertical perturbation would be technologically easier, since the response sensitivity is of higher order.

### 3. Experimental validation of the ASFS waveguide

In order to corroborate the analytical and numerical results, an ASFS waveguide prototype was fabricated using 3D-printing with a poly-lactic acid (PLA) filament employing a truss-like substructure for the solid parts of the structure of the waveguide. The geometrical parameters of the fabricated structure are $d$ = 34.3 mm, $a_1 = a_3 = g$ = 10.3 mm, $h_1$ = 20.6 mm, $h_3$ = 18.2 mm, and the width in the $z$ direction is equal to 20 mm, while the fluid is air with the parameters $\rho$ = 1.1802 $kg/m^3$, $C$ = 346.87 $m/s$. Due to the need for a sufficient number of unit cells and, at the same time, the limitations of the 3D printer, three equal segments were fabricated separately, each comprised of 7 unit cells, as shown in Figure 5.

The three segments were attached to an aluminum profile and tightened for straight alignment and pressed against a commercial speaker attached to a sound blaster device connected to a laptop. Two small microphones were mounted with a distance $d$ between them on the central segment, with the head of each microphone sunk into a drilled hole in the waveguide, allowing the microphone to perform as a probe of pressure inside the upper corrugation of the waveguide. A slow chirp signal with linear time-increase in frequency was created using the audio-analysis software Audacity [76] and played using the speaker. Consequently, the phase shift between the two microphones was measured and recorded.

The relative phase shifts and the corresponding applied excitation frequencies were used to obtain experimental dispersion diagrams. For the purpose of a better comparison of the simulated and measured results, a 3D model was created in COMSOL, and the corresponding dispersion curves were calculated. The comparison in Figure 6 reveals an excellent agreement between simulated and measured results.

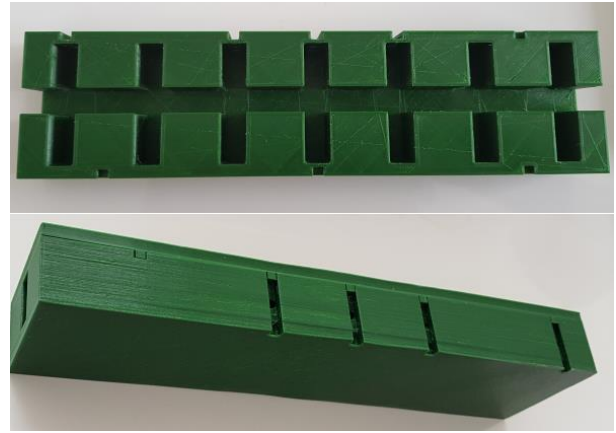

(a)

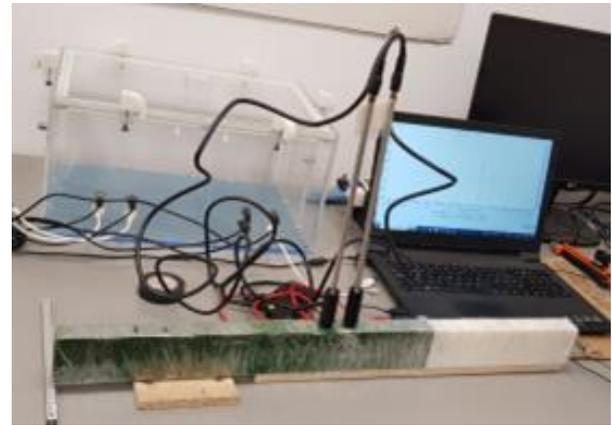

(b)

Figure 5: (a) Fabricated segment of the ASFS waveguide, (b) Measurement setup.

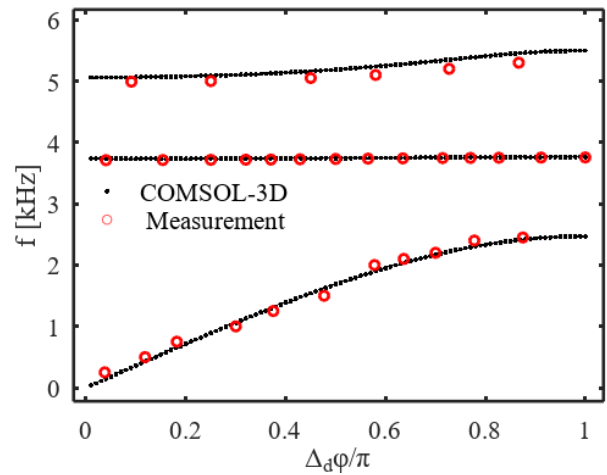

Figure 6: Comparison of calculated and measured dispersion curves for the ASFS waveguide.

## 4. A $CO_2$ sensor based on the ASFS waveguide

As discussed above, acoustic gas sensors typically rely on the fundamental modes of the acoustic structures, while the sensing principle is predominantly based on the amplitude change of a signal. As it happens, however, amplitude measurements are typically prone to be corrupted by noise, although this can be mitigated to some extent [44]. This is one of the consequences of the fact that low-amplitude pressure measurement devices (microphones) are normally relative rather than absolute and can be affected by surrounding sounds of different frequencies. The amplitude of resonance in a higher mode is usually lower than in the fundamental mode, and hence it is rarely used for amplitude measurements. While there are sensors based on the measurement of the speed of sound [40-43], which is embedded also in the phase measurement principle, those sensors rely on fundamental modes, which limits their sensitivity. Namely, fundamental modes have a linear dispersion curve for most wavenumbers, which in principle challenges their utilization in sensitive large-range sensors. In contrast, due to their slow-wave property, nearly-flat-dispersion modes can be an excellent choice for high sensitivity sensing that is based on measuring phase changes. In this context, we demonstrate a novel sensor employing the second mode of an ASFS waveguide and a phase-measurement principle, aimed at the detection of the concentration of $CO_2$ in composition of air in the vicinity of the sensor.

Figure 7(a) shows the numerically obtained dispersion curves for an ASFS waveguide for various $CO_2$ concentrations. The geometric parameters were optimized such that the second mode appears around 11 kHz and their values are: $d$ = 11.5 mm, $a_1$ = $a_3$ = $g$ = 3.5 mm, $h_1$ = 6 mm, $h_3$ = 6.9 mm, and the width in the $z$ direction is equal to 12.5 mm. The geometric-parameters values were decreased compared to the previously shown ASFS waveguide to correspond to a more compact structure and to match our fabrication and measurement capabilities, as detailed in the following section. The change in the $CO_2$ concentration was modeled in simulations through its effect on the values of the density $\rho$ and speed of sound $C$ of a mixture of the air and $CO_2$, using the following formulas:

$$\rho = \rho_{air}^{23\,°C}\left[1+\left(\frac{M_{CO_2}}{M_a}-1\right)c_v\right], C= \sqrt{\frac{1+2[(1-c_v)\alpha_{air}+c_v\alpha_{CO_2}]^{-1}}{1+\frac{2}{\alpha_{air}}}}\frac{C_{air}^{23\,°C}}{\sqrt{1+\left(\frac{M_{CO_2}}{M_a}-1\right)c_v}}, \quad (1)$$

$$\frac{M_{CO_2}}{M_a}=\frac{44.01}{28.9647}, \alpha_{CO_2}=6, \alpha_{air}=5.003, \rho_{air}^{23\,°C}= 1.192\frac{kg}{m^3}, C_{air}^{23\,°C}=344.96\frac{m}{s}. \quad (2)$$

where $M_x$ denotes molecular weight of material $x$, and $\alpha$ is the number of degrees of freedom of a gas molecule.

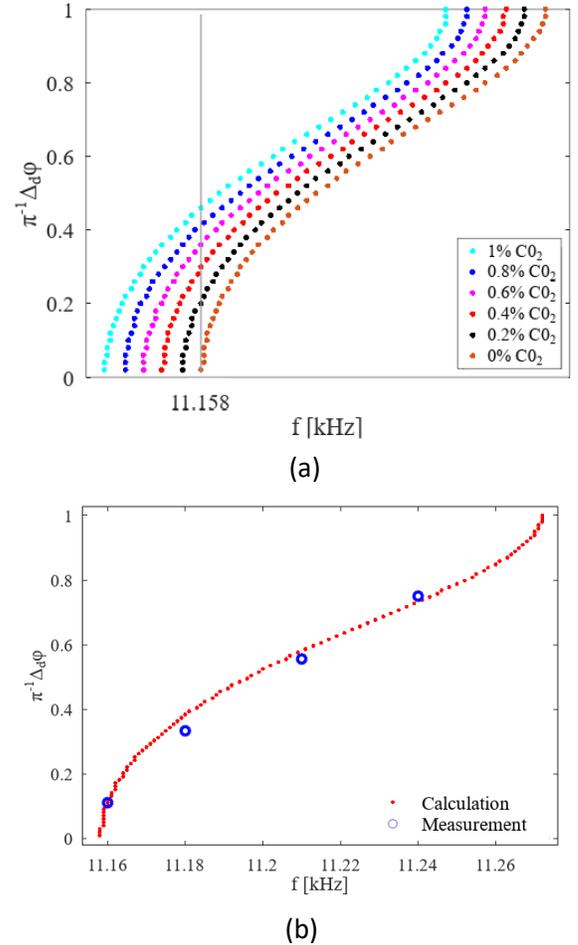

Figure 7: Calculated dispersion diagrams of the second mode of an ASFS waveguide for various concentrations of $CO_2$ (a) and validation of 0% $CO_2$ case against measurement (b).

We note that CO$_2$ concentration of 0% means that the mixture consists of pure air, which includes 0.04% of CO$_2$. By the same token, CO$_2$ concentration of 1% means that there is 1.04% of CO$_2$ in the mixture, i.e., 1% of the CO$_2$ has been added.

One observes that the different dispersion curves have the same shape yet there is a slight red spectral shift as the CO$_2$ concentration increases. The grey vertical line going through different-concentration dispersion curves for the frequency of 11.158 kHz shows how different concentrations cause different phase accumulations along the waveguide. It should be noted that the structure has to be cautiously optimized so that the second mode becomes only nearly flat (and not perfectly flat). Then, the mode would not represent a standing-wave excited at an isolated frequency value, but rather would enable slow wave propagation and allow the measurement of the phase-change along the waveguide. Moreover, propagation would occur at a specific frequency for a whole range of values of a target analyte. As a side note, in the case of a *perfectly*-flat band, infinitely many wavenumber values would correspond to the excitation frequency and we would not be able to measure the wavenumber.

According to the performed numerical simulations, the proposed ASFS waveguide has a sensitivity of 0.003 Hz/ppm, which means that a concentration change of 300 ppm produces a 1 Hz shift in the frequency response. However, one should not be misled by sensitivity value since our sensor is based on the phase-change response to the concentration change, and in such a scenario the full phase range is adjusted to the prescribed range of CO$_2$ concentrations, while the frequency range is a consequence of the chosen concentrations range. Therefore, the overall sensitivity is equal to $\pi/\Delta c_v$, so it is smaller for a larger range of concentrations. Furthermore, the sensor utilizes nonlinear phase response, which allows for higher phase response to small concentration changes specifically at low concentrations. There, this sensitivity is not expressed in a high value of Hz per ppm, since we do not suggest measuring the frequency shift itself.

Instead, we suggest measuring the phase shift, and indeed, we have a large value of phase shift per ppm of concentration for low concentration values. This high sensitivity is not manifested in a high value of a slope of a linear curve, but rather in the fact that the response is concave, and for low concentrations the (variable) slope becomes very high. This concave response is achieved by adjusting the operational frequency to have $\frac{\partial \omega}{\partial k} = \frac{\partial f}{\partial \Delta_d \varphi} = 0$ (similarly to the approach of [77-78]) at the zero-phase limit.

A practical issue related to the resolution of the sensor is the following. In COMSOL calculations at frequencies around 10 kHz, the accuracy of the eigenvalue solver is $10^{-4}$, which means 1 Hz frequency resolution, or 300 ppm concentration resolution. Also, for the actual sensor, in a measurement, the absolute sensitivity is lower: the maximum sampling frequency of a standard sound card is around 34 times the operational frequency chosen here, which implies 17 sampled points per half-period, or 1 sample point for 500 ppm. Therefore, it may be stated that the maximum possible absolute resolution of the sensor with standard electronic equipment would be around 500 ppm. The normal concentration of carbon dioxide in air is 400 ppm, which means that the design sensor is sufficient for detecting concentrations from the perspective of health hazards in inhalation.

Figure 7(b) shows open-air measurements with the sensor at various frequencies in the relevant range covering all the relevant phase range. One observes very good experimental validation of the calculations.

In the following section, the sensing principle is demonstrated experimentally, and the sensor's properties and performance are discussed in detail.

5. **Experimental testing of the CO$_2$ sensor**

To test the proposed sensor experimentally, the structure was fabricated using 3D printing technology, employing a PLA filament, and the acoustic response was measured for several relevant

$CO_2$ concentration values in controlled environment. Prior to fabrication and measurements, several issues had to be resolved. Specifically, the dimensions and the quantity of the unit cells in the ASFS waveguide sensor were determined by the 3D-printer resolution (dictating the minimal geometric dimension in the waveguide), the size of the available microphones and size of the gas chamber available for the experiment; the necessity to have a sufficient number of unit cells to achieve a well-resolved wave structure and, ultimately, the desire to have a more compact structure compared to the previously presented ASFS waveguide.

Several 3D simulations were run to determine the final configuration and geometric parameters, where the analyzed structures were comprised not only of the ASFS waveguide but also of holes for the microphones, which were modeled using radiative boundary conditions.

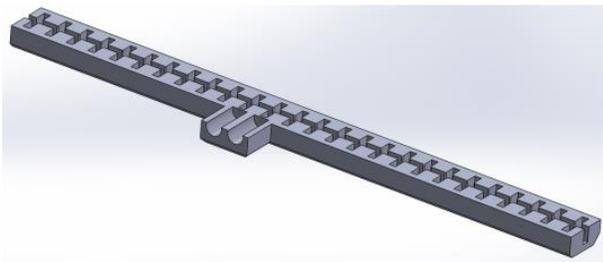

(a)

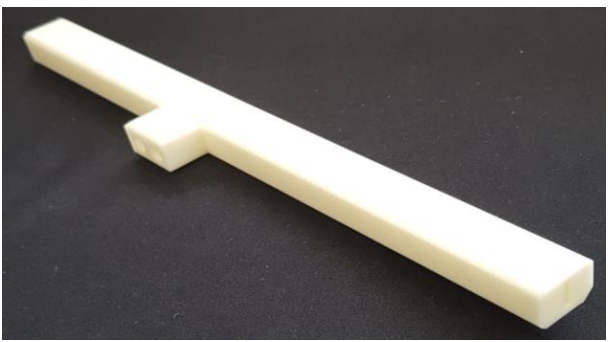

(b)

Figure 8: (a) Cross section view of the numerically analyzed sensor, (b) Isometric view of the fabricated sensor

Dispersion diagrams for the optimized waveguide structure for different $CO_2$ concentrations were obtained following the principle explained in the previous section and using Eqs. (1) and (2), where the temperature was measured and set to 23 °C. Since the calculated dispersion diagrams turned out to be practically the same as for the initial structure, we will omit them here. The operating frequency was chosen to be 11.158 kHz. This value produces maximum measured-phase sensitivity at small concentrations and enables covering the full range of relevant $CO_2$ concentrations.

The number of unit cells sufficient to support a wave with a nearly flat dispersion curve was found to be 26 unit cells, with 15 unit cells between the speaker and the first of the two 8 mm-diameter microphones, and 11 unit cells between the second microphone and the outlet. A larger number of unit-cells would make the sensor too large for the gas chamber, and is essentially not necessary. Isometric views of the numerically analyzed and fabricated structure are shown in Figure 8.

The extrusion depth for the 2D design was set to be 12.5 mm and the thickness of the plates representing the hypothetically rigid corrugated surface was set to 3.2 mm. The fabricated upper corrugations corresponding to the locations of the two microphones were covered by 1 mm-thick plates, with a 1 mm-diameter central hole for the sound to get to the head of the microphone positioned against the said 1 mm ceiling. Two microphones constructed as cylinders with flat sound-receptive circular heads, were positioned perpendicularly to the axis of the waveguide, at adjacent unit cells, centered above the vertical corrugations. The cylinders were mounted using plastic nests printed monolithically with the structure.

To perform acoustic measurements, two microphones on cylindrical sticks were inserted into the sockets of the sensor, as depicted in Figure 8, having the air pressure wave reaching them through the 1 mm-diameter hole. A small speaker of an earphone was attached by duct tape to one end of the waveguide, 15 unit-cells distant from the microphone sockets, with the other waveguide end remaining open. The sound was generated using the free software Audacity. Measurements were conducted in a laboratory, using a gas chamber, into which $CO_2$ was pumped through a 1 mm-diameter tube from a $CO_2$ balloon, with 0.1 bar relative pressure during the course of approximately 1 minute, until reaching a concentration higher than 10000 ppm, according to a reference commercial optical IR $CO_2$-concentration sensor ("Qingping Air Monitor Lite"). Subsequently, a manual air pump was used to pump in air and let the $CO_2$-polluted air flow out through another 1 mm-diameter tube until desired concentrations were reached, as confirmed by the reference sensor. Acoustic measurements were then performed at steady-state concentrations of 0.7%, 0.5%, 0.3% and 0.1%, in addition to the 0% concentration measurement that had been conducted outside the gas chamber. The experimental setup is shown in Figure 9.

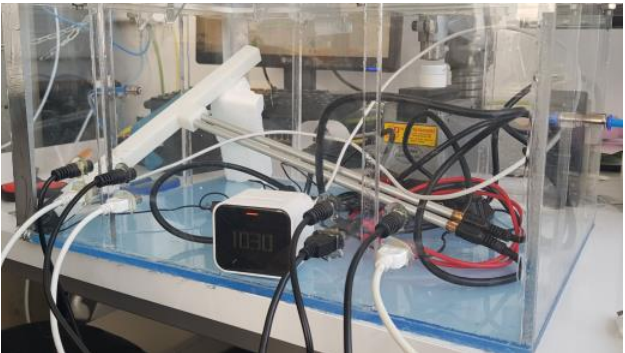

Figure 9: Experimental setup: a PMMA box (with an inlet for gas and an outlet) containing the fabricated sensor, with two microphones inserted in the sockets, an earphone speaker attached by duct tape at the longer end, a soundblaster device Omni Surround 5.1, connecting the three audio devices to the computer outside the chamber using USB ports installed in the chamber, with a piece of polystyrene to support the sensor to have a free outflux, and a reference sensor Qingping Air Monitor Lite.

The measured phases obtained using a procedure discussed in detail in Appendix B, are depicted on top of the calculated values in Figure 10. The concentration for each measurement was read from the reference Qingping Air Monitor. It should be noted that the reference sensor measures total $CO_2$ concentration using an optical absorption mechanism. Due to this, the monitor of the reference sensor shows the sum of $CO_2$ in standard air and the additionally pumped $CO_2$. To deduce the additionally pumped $CO_2$ concentration, an amount of 0.04% had to be subtracted. The reason for this is that the computation involved specifying the speed of sound of standard air at a certain temperature, and standard air already contains 0.04% $CO_2$.

One can observe excellent correspondence between the measured and calculated concentrations, i.e., between the concentrations measured by the reference sensor and those measured by the proposed sensor. The sensor shows high accuracy already at volume concentrations as small as 0.04% $CO_2$, which is a non-hazardous concentration, while the steep initial rise guarantees good phase resolution even for a small concentration change.

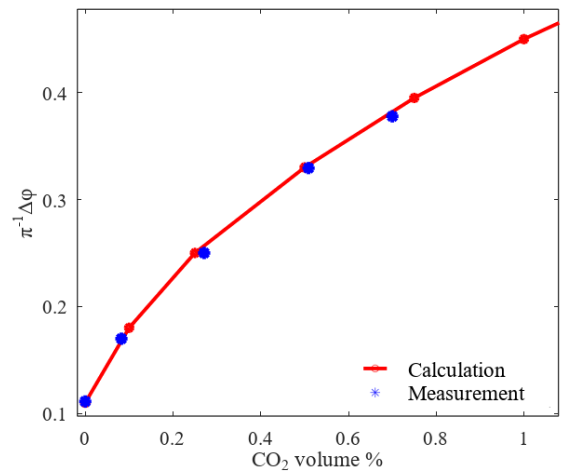

Figure 10: Comparison of the calculated and measured relative phase change versus the concentration of $CO_2$.

## 6. Discussion

The proposed sensor represents a novel sensing platform characterized by high accuracy and sensitivity, low-cost, and the possibility for a rapid and reliable measurement with low maintenance.

We first implemented a design that adds a perturbation to an otherwise symmetric waveguide, characterized by a perfectly-flat second band. The symmetry is broken vertically, in a fashion that maintains the uniformity of the waveguide in the propagation direction, while achieving a nearly-flat mode, similarly to the idea in [62]. This is different from some other designs that introduce horizontal perturbation to achieve a nearly-flat mode [65]-[66]. If the horizontal perturbation is breaking the periodicity of the waveguide, then measuring the wavenumber of propagating waves will be problematic. If the horizontal perturbation preserves the periodicity, still horizontal symmetry-breaking enforces fabrication requirements in the propagation direction. When approaching the manufacturing limit for a small-enough perturbation (for high-enough sensitivity still enabling measurement), the fabrication inaccuracy may destroy the propagating wave, damaging the ability to measure a phase difference as an indication for the concentration, with its advantages in terms of robustness to noise and efficient measurement time. A vertical perturbation helps to overcome such difficulties. The absolute size of the vertical asymmetry can be adjusted to the fabrication capabilities and the relative accuracy can be decreased by increasing the average height of the vertical perturbation, which has the only effect of decreasing the operational frequency, which is advantageous anyway, as it increases the accuracy of signa processing for available equipment.

The most important aspect of the introduced asymmetry is that it provides the possibility to change the nature of the second mode from perfectly-flat to nearly-flat. While the proposed ASFS structure and its analysis are presented for the first time, the major novelty in this work is the sensor and its sensing principle that is based on a nearly-flat mode, also tailored to maximum sensitivity, in an ASFS waveguide, and its reliance on phase difference measurement.

The significance of the proposed sensing principle is twofold. The majority of the proposed sensors are based on damped-resonance amplitude-change, rather than on phase-measurement, and unlike phase measurement, amplitude measurement is prone to modulation and attenuation by noise and measurement duration. This problem was addressed in [44], but only partially. Therefore, the phase difference measurements are significantly more robust and as such represent a more reliable solution.

While there are in the literature sensors based on the measurement of the speed of sound [40-43], which is embedded also in the phase-measurement principle, those sensors employ the fundamental mode, and this is true for most acoustic sensors. The major disadvantage of the fundamental mode is their linearity in the majority of the wavenumber range, which limits the sensitivity, i.e., they cannot provide sufficiently different wavenumbers for minute changes in the fluid as does the nearly-flat mode.

Furthermore, when introducing a small vertical perturbation to the symmetric design, the second-mode band ceases to be flat for all wavenumbers but remains flat for the zero wavenumber. If the operational frequency is chosen in such a way that the wavenumber of the second-mode for standard air is equal to zero then the measured wavenumber increases sharply even for infinitesimally-small changes in gas concentration, which provides a highly sensitive phase–concentration response, allowing very high sensing resolution for small concentrations. Therefore, the second advantage of our solution is the employment of a nearly-flat dispersion band.

As previously discussed, our sensitivity is equal to 0.003 Hz/ppm, however, this value should be taken cautiously and interpreted in the context of the given sensing principle. Since the robustly measurable value is phase, we design the sensor such that the phase difference between adjacent unit-cells spans $\pi/2$ radians for the relevant $CO_2$ concentrations range. Also, the sensor is designed such that the phase-concentration response curve is nonlinear, resembling a square-root function, for which the sensitivity, or the slope, becomes arbitrarily high at low concentrations. Therefore, the smaller the concentrations range, the higher is the sensitivity value, i.e., the sensitivity can be chosen almost arbitrarily. High -sensitivity values are limited only by practical considerations, such as the accuracy of the numerical algorithm of the employed software and the sampling rate of the used electronic equipment.

The sensitivity of periodic-structure-based sensors, as those in Refs. [53,61], are of the order of 10mHz/ppm, which might be considered similar to that of the proposed one, but one should note that these sensors employ almost-linear modes and operate at frequencies of around 60 kHz, which is 5.4 times higher than the proposed operating frequency. Quartz-Crystal-Microbalance (QCM), SAW or BAW, and Film-Bulk-Acoustic-Resonator (FBAR) sensors typically exhibit sensitivity of tens of Hz/ppm, $10^{-3}$–$10^{-2}$ kHz/ppm, and around 1kHz/ppm, respectively, while their operating frequencies are in the range of 1–10 kHz, 100 kHz – 100 GHz, and 100 MHz – 20 GHz, respectively [49]. Therefore, their sensitivity can be considered comparable to that of the proposed sensor. As for the limit of detection, it is indeed smaller in other acoustic sensors; however, those sensors are typically aimed at other applications, such as routine air-quality monitoring.

Besides novelty, increased sensitivity, and robustness to noise, the proposed sensor shows some other advantages in comparison to other acoustic sensors. It was shown that even for small concentrations the proposed sensor is very accurate. Although only measurements up to 1% concentration have been shown, it should be noted that the proposed sensor can measure concentrations up to 3% or even up to 5% with a slight change in the sensor design. In other words, the proposed sensor can work in a large range of concentrations, without reaching saturation and as such be applied in various industrial applications.

In comparison to other acoustic sensors, which typically require advanced fabrication processes, the proposed sensor can be printed from cheap PLA material and requires only an earphone and two microphones, the signals from which can be coordinated by a simple electronic board and analyzed using a standard microprocessor. In addition to these advanced properties, the sensor has shorter measurement time in comparison to others, it is label-free, provides rapid and robust response, and does not require relaxation time or air flushing. Consequently, it can be used for real-time measurements.

Since the basic detection mechanism is related to molecular weight of a gas, the sensor is equivalently operative for detection of any heavy gas, such as $N_2O$, $C_3H_8$, $C_4H_{10}$ or light gas such as $CH_4$ or $NH_3$. At the same time, this represents an obstacle for the sensor's selectivity since several gases can have (approximately) the same molecular weight.

The size of the proposed sensor may represent a challenge. While not all applications require small sensors, miniaturization is an important requirement in sensor design. We would like to note that a significantly smaller version of the sensor can be achieved if the lateral dimensions are modulated, however, such a configuration requires more advanced 3D printing technology with higher printing resolution. Our future work will include the improvement of the structure and its miniaturization using more advanced technologies.

## 7. Conclusions

In this work, we have investigated an ASFS waveguide and its application in gas sensing. First, a numerical method was suggested for robust dispersion analysis based on the expansion of the acoustic fields in the direction perpendicular to the wave propagation direction, using a basis of wave functions, but employing a point-collocation rather than mode-matching technique. The method showed satisfactory results coinciding with finite element analysis implemented in the commercial software COMSOL-Multiphysics. An ASFS waveguide was fabricated using 3D printing technology and its responses were measured in a simple measurement setup that includes off-the-shelf microphones and a speaker. The measurement results showed excellent agreement with the calculated ones.

Furthermore, it was numerically demonstrated how the second mode of an ASFS waveguide that has nearly-flat nature can be used for gas sensing purposes. Consequently, a numerically-optimized $CO_2$ sensor was fabricated using 3D printing technology and experimentally tested in a gas chamber-controlled $CO_2$ concentration. The measured results matched well to the calculated ones, demonstrating the potential of the proposed sensor. Besides the novelty of the sensing principle, the sensor is characterized by high sensitivity, accuracy, low costs, robustness to noises, rapid response without the need for relaxation time and maintenance, and the possibility to operate in a large range of concentrations without reaching saturation. Consequently, we opt that the proposed work represents a valuable contribution not only to the theoretical analysis of acoustic periodic structures but also to sensor engineering applicable for real-world problems.

**Acknowledgements**

The work described in this paper was conducted within the project NOCTURNO, which received funding from the European Union's Horizon 2020 research and innovation programme under Grant No. 777714. This work was also supported through the ANTARES project that has received funding from the European Union's Horizon 2020 research and innovation programme under grant agreement SGA-CSA No. 739570 under FPA No. 664387, https://doi.org/10.3030/739570.

The work was also supported by the Department of Defense, the Air Force Office of Scientific Research, and the Simons Foundation.

**Appendix A: Derivation of the calculated dispersion diagram for the asymmetric spoof-fluid-spoof waveguide**

The basic governing equations of a fluid subjected to acoustic excitation are as follows:

$$\rho\dot{\boldsymbol{v}} = -\nabla p, \dot{\rho} + \rho \nabla \cdot \boldsymbol{v} = 0, \frac{\partial p}{\partial \rho} = C^2, \nabla \times \boldsymbol{v} = 0 \Rightarrow \nabla^2 p = \frac{1}{C^2}\frac{\partial^2 p}{\partial t^2}, \nabla^2 \boldsymbol{v} = \frac{1}{C^2}\frac{\partial^2 \boldsymbol{v}}{\partial t^2}. \quad (A.1)$$

In frequency space the following relations hold:

$$\nabla^2 \breve{p} + k^2 \breve{p} = 0, \nabla^2 \breve{\boldsymbol{q}} + k^2 \breve{\boldsymbol{q}} = 0, \nabla \breve{p} = k \breve{\boldsymbol{q}}, \nabla \cdot \breve{\boldsymbol{q}} = k \breve{p}, \quad (A.2)$$

where

$$\hat{p} = e^{-i\omega t}p, \hat{\boldsymbol{v}} = e^{-i\omega t}\boldsymbol{v}, k = \frac{\omega}{c}, \eta = \rho c, \hat{\boldsymbol{q}} = \eta \hat{\boldsymbol{v}}, \hat{\boldsymbol{q}} = -i\breve{\boldsymbol{q}}, \hat{p} = -i\breve{p}, \quad (A.3)$$

and where ρ, p, C, $\boldsymbol{v}$, ω are the density, pressure, speed of sound, material velocity and excitation frequency in the fluid, respectively.

It is next assumed that the excitation is uniform in the z direction, and consequently, so is the response. Therefore, the following equations are assumed to hold in the x-y plane:

$$\frac{\partial^2 \breve{p}}{\partial^2 x^2} + \frac{\partial^2 \breve{p}}{\partial^2 y^2} + k^2 \breve{p} = 0, \breve{q}_x = k^{-1}\frac{\partial \breve{p}}{\partial x}, \breve{q}_y = k^{-1}\frac{\partial \breve{p}}{\partial y}. \quad (A.4)$$

The general solution of the Eq. (A.4) is

$$\breve{p}(x,y,z) = \breve{p}(x,y) = \sum_m \left[\left(A_m e^{-ik_{x,m}x} + B_m e^{ik_{x,m}x}\right)e^{-i\sqrt{k^2-k_{x,m}^2}y} + \left(C_m e^{-ik_{x,m}x} + D_m e^{ik_{x,m}x}\right)e^{i\sqrt{k^2-k_{x,m}^2}y}\right] \quad (A.5)$$

A wave solution (with $k_{x,m} \in \mathbb{R}$) would be supported for waves with small enough wavelength, namely, $k_{x,m} \gg x_{max}^{-1}$. It is further assumed that, for any value of z, the fluid occupies in the x-y plane some subspace bounded by straight-segment boundaries as depicted in Figure 1.

One can then specify the coefficients $A_m$, $B_m$, $C_m$, $D_m$, $k_x$ in each of the three sub-domains in a unit cell of horizontal length d. In the domains I and III the condition of vanishing normal velocity on each of the three rigid boundaries produces the following solutions:

$$\breve{p}^{III}(x,y) = \frac{1}{a_3}\sum_{m=0}^{\infty} A_m^{III} \frac{\cos(m\pi x/a_3)\cos\left[\sqrt{k^2-(m\pi/a_3)^2}\left(\frac{g}{2}+h_3-y\right)\right]}{\sqrt{k^2-(m\pi/a_3)^2}\cos\left(\sqrt{k^2-(m\pi/a_3)^2}h_3\right)}, \quad (A.6)$$

where the coefficients $A^{III}_m$ are to be determined by requiring that $\breve{p}^{III}(x,g/2) = \breve{p}^{II}(x,g/2)$.

$$\check{p}^I(x,y) =$$
$$\frac{1}{a_1}\sum_{m=0}^{\infty} A_m^I \frac{\cos(m\pi x/a_1)\cos\left[\sqrt{k^2-(m\pi/a_1)^2}\left(\frac{g}{2}+h_1+y\right)\right]}{\sqrt{k^2-(m\pi/a_1)^2}\cos\left(\sqrt{k^2-(m\pi/a_1)^2}h_1\right)},$$
(A.7)

and the coefficients $A'_m$ are to be determined by requiring that $\check{p}^I(x,-g/2) = \check{p}^{II}(x,-g/2)$.

The horizontal boundary conditions in domain I and III impose $a_{1,3}$-periodicity on the solution, since the normal velocities have to be equal and vanishing at both $x = 0$ and $x = a_{1,3}$. In contrast, in the middle layer there are no horizontal boundary conditions, and thus for a given wave with its given wavelength, the phase difference between $x = 0$ and $x = d$ would generally be non-vanishing. Hence, the solution in domain II will not be $d$-periodic in $x$, but at best $d$-periodic up to a phase shift.

For every value of $y$, for $x$-periodic waveguides, the spectral wave-equation is linear second-order in $x$ with no first derivative and with periodic coefficients, which makes it a Hill equation. For the Hill equation a Floquet-theory result exists [79] according to which the solution is $d$-periodic up to a phase shift, namely, that $dk_{x,m} = 2m\pi + \Delta_d\phi, m \in \mathbb{Z}$. Thus, one has

$$\check{p}^{II}(x,y) = \frac{1}{d}\sum_{m=-\infty}^{\infty} \frac{e^{-i\left(m+\frac{\Delta_d\phi}{2\pi}\right)\frac{2\pi}{d}x}}{\sqrt{k^2-\left[\left(m+\frac{\Delta_d\phi}{2\pi}\right)\frac{2\pi}{d}\right]^2}} \times$$
$$\left\{A_m^{II} e^{-i\sqrt{k^2-\left[\left(m+\frac{\Delta_d\phi}{2\pi}\right)\frac{2\pi}{d}\right]^2}y} + B_m^{II} e^{i\sqrt{k^2-\left[\left(m+\frac{\Delta_d\phi}{2\pi}\right)\frac{2\pi}{d}\right]^2}y}\right\},$$
(A.8)

where $\Delta_d\phi \in [0,2\pi)$ is the phase shift over distance $d$ of a wave with a given frequency $\omega$. The coefficients $A_{\pm|m|}^{II}, B_{\pm|m|}^{II}$ would be determined by the requirement to comply with the vanishing of $\check{q}_y$ at the vertical boundaries at $a_3 < x < d$, $y = g/2$ and $a_1 < x < d$, $y = -g/2$, and with continuity of normal velocities, namely, $\eta^{-1}\check{q}_y$ at $0 < x < a_3$, $y = g/2$ and $0 < x < a_1$, $y = -g/2$.

If the waveguide is comprised of $N_d$ unit cells, then it would support wave solutions with wavelength satisfying the condition $\Delta_d\phi \gg N_d^{-1}$. For example, for $N_d = 15$, adopting that $10 \gg 1$, one could have waves for $\frac{\Delta_d\phi}{\pi} > 0.2$. Due to the symmetry with respect to the sign of $x$, the relevant range of normalized wavenumbers for the said example would be $\frac{\Delta_d\phi}{\pi} \in [0.2,1]$. The said symmetry would also render the frequency dependence on wavenumber very low for $\frac{\Delta_d\phi}{\pi} > 0.8$. Consequently, one would be able to observe waves in the wavelength range of $\frac{\Delta_d\phi}{\pi} \in [0.2,0.8]$. The smallest possible wavelength in a $d$-periodic waveguide is $d$, so shorter waves will not propagate. The largest wavelength for propagating waves would be, as aforesaid, of wavelength around $0.1N_dd$. Consequently, a dispersion study is only possible for $N_d > 10$.

For the possibility of a numerical solution of the collocation type, the $x$ coordinate in the range (0,$d$] is divided as uniformly as possible into segments, or rather a set of discrete points is chosen in the said range, with as uniform distances as possible between them, to enforce the boundary conditions. More specifically, at $y = g/2$, the segment $0 < x < a_3$ is divided into $n_{a_3}$ uniformly spaced points, and the segment $a_3 < x < d$ is divided into $n_{s_3}$ uniformly spaced points. Similarly, at $y = -g/2$, the segment $0 < x < a_1$ is divided into $n_{a_1}$ uniformly spaced points, and the segment $a_1 < x < d$ is divided into $n_{s_1}$ uniformly spaced points.

The stability of this spectral collocation method is guaranteed for the following choice of parameters:

$$n_{s_{1,3}} = \begin{cases} \left\lceil\frac{d-a_{1,3}}{a_{1,3}}n_{a_{1,3}}\right\rceil, & \text{if } \left\lceil\frac{d-a_{1,3}}{a_{1,3}}n_{a_{1,3}}\right\rceil \text{ is even} \\ \left\lceil\frac{d-a_{1,3}}{a_{1,3}}n_{a_{1,3}}\right\rceil + 1, & \text{if } \left\lceil\frac{d-a_{1,3}}{a_{1,3}}n_{a_{1,3}}\right\rceil \text{ is odd} \end{cases}$$
(A.9)

where $\lceil\cdot\rceil$ is the ceiling function, or the closest larger integer number.

In addition, the infinite series have to be truncated in numerical calculations. The upper limit of infinity in the given solutions is hence replaced by $N_I$, $N_{II}$, and

$N_{III}$. In order to extract unambiguously the $N_{III} + 1$ coefficients $A_m^{I,II}$ from equating the pressure in domains I and III at the $n_{a_{1,3}}$ collocation points to the pressure at domain II at the same points, the following relations must hold:

$$n_{a_{1,3}} = N_{I,III} + 1. \quad (A.10)$$

Finally, for the number of equations to correspond to the number of unknown coefficients, the following relation must hold:

$$N_{II} = \frac{1}{4}\left(2n_{a_1} + 2n_{a_3} + n_{s_1} + n_{s_3} - N_I - N_{III} - 4\right). \quad (A.11)$$

Furthermore, symmetry between the lower and upper boundaries requires that $n_{a_1} + n_{s_1} = n_{a_3} + n_{s_3}$. Consequently, if this requirement is satisfied, the following relation would hold: $2N_{II} = N_I + n_{s_1} = N_{III} + n_{s_3}$. Since $n_{s_3}$ is even, a necessary condition is that $N_I$ be even, and that $N_{III}$ would satisfy the relation

$$N_{III} + n_{s_3} = N_I + n_{s_1}. \quad (A.12)$$

This means that there is only one free parameter in this algorithm, namely, $N_I$, which can be an arbitrary even integer. The chosen value should be large enough with respect to unity to obtain convergence, but not too large to avoid resonance with numerical noise. In the solved examples the choice $N_I$ = 10 was made.

An important remark should be made here. The differential equation (A.4) is smooth in each smooth-boundary domain, and thus its solution is smooth. Therefore, the Fourier coefficients of the solution should decay exponentially in domains I and III and they are proportional to $A_m^I, A_m^{III}$. In domain II, the solution is a Fourier series in *x*, multiplied by a low frequency oscillatory function of *x*. Consequently, the solution would not contain highly oscillatory large amplitude modes. Thus, enforcing continuity at a large number of collocation points will guarantee *small* field discontinuities between the collocation points, since the solutions would not be able to oscillate with a large amplitude between them. Therefore, in this case, the collocation method of enforcing continuity at interfaces is expected to be reliable and convergent.

Writing down the continuity equations for pressure and vertical velocity at 0 < *x* < $a_3$, *y* = *g*/2 and 0 < *x* < $a_1$, *y* = -*g*/2 and the condition of vanishing vertical velocities at $a_3$ < *x* < *d*, *y* = *g*/2 and $a_1$ < *x* < *d*, *y* = -*g*/2 one obtains a system of 6 sets of equations for 6 sets of variables ($\mathbf{w} = \{A_m^I, A_{\pm|m|}^{II}, B_{\pm|m|}^{II}, A_m^{III}\}^\top$). Obtaining the representative matrix $\mathbf{H}$ of this system of equations, namely, $\mathbf{Hw} = 0$, requires only the substitution of a set of uniformly spaced points in each segment *x*, namely,

$$x \to \{x\}_j, x_j^{(a_1,a_3)} = \frac{a_{1,3}}{n_{a_{1,3}}+1}j, j = 1,2,\ldots,n_{a_{1,3}}, x_j^{(s_1,s_3)} = a_{1,3} + \frac{d-a_{1,3}}{n_{a_{1,3}}+1}j, j = 1,2,\ldots,n_{a_{1,3}}. \quad (A.13)$$

After the matrix $\mathbf{H}$ is constructed, with each column corresponding to a mode of the solution in the entire domain and each row corresponding to an equation on a collocation point, numerical eigenvalue analysis of $\mathbf{H}$ is performed and the eigenvalue with the minimum absolute value is identified.

Nonzero-amplitude pressure waves can be supported by the studied system only for a vanishing minimum absolute-value eigenvalue of $\mathbf{H}$. The logarithm of this measure is plotted against the wave frequency for $\Delta_d \phi = \pi/2$ for the following geometrical parameters *d* = 34.3 mm, $a_1$ = $a_3$ = *g* = 10.3 mm, $h_1$ = 20.6 mm, $h_3$ = 18.2 mm, *C* = 346.87 m/s, $\rho$ = 1.1802 kg/m³ in Figure A.1.

For high enough frequency resolution and spatial resolution, the narrow dips as shown in Figure A.1 would approach minus infinity and the corresponding frequencies would be the nonzero-amplitude wave frequencies for the assumed value of $\Delta_d \phi$.

In order to obtain the wavenumber for a given frequency, one has to measure the phases of the propagating pressure waves at two locations in the waveguide with a distance *d* between them.

If one acknowledges a relative error $\delta$ in the distance between two probes measuring the pressure,

rendering the actual said distance to be (1+δ)d, then modes corresponding to terms with $m = \mathcal{O}(\lfloor \delta^{-1} \rfloor)$ may produce additional parasitic finite phase shift.

However, for such terms the denominator in Eq. (A.8) would be of the order of $m = \mathcal{O}(\lfloor \delta^{-1} \rfloor) \gg 1$, rendering the contribution of the term insignificant.

The latter would hold due to the fact that the coefficients $A_m^{II}, B_m^{II}$ decay exponentially for $m \gg 1$, balancing the braced expression to order of unity for y = g/2. This analysis shows that the pressure is stable with respect to phase-shift measurement at the waveguide boundaries. The coefficients may decrease even more strongly, but not more weakly, as the alternating harmonic series is already marginally convergent.

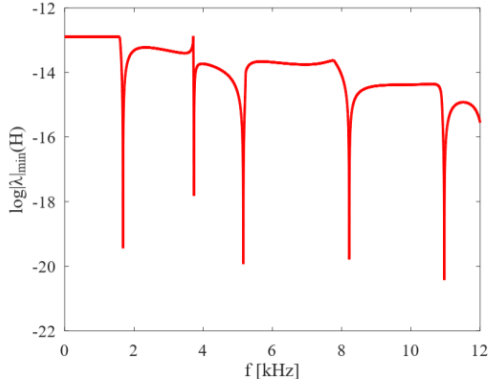

Figure A.1: Plot of $\log|\lambda|_{min}(\mathbf{H})$ vs. $f = \frac{\omega}{2\pi}$.

**Appendix B: Derivation of the measured dispersion diagram**

The procedure for the derivation of the measured dispersion diagram goes as follows: two microphones are positioned in the middle of the structure one unit-cell apart and record a signal approaching from the speaker positioned at end of the waveguide. The recorded pressure-level signals are shown in Fig. B.1, first in the full recording range, then in the range of 10 msec from 4.102sec to 4.112 sec and finally over the range of 0.2msec, from 4.1104sec to 4.1106sec.

The phase difference between the signals should be checked at the right time, which is when the wavefront have reached the open end of the waveguide and the rarefaction wave has returned to the microphones.

$$\Delta t_\phi = \frac{\Delta L}{\frac{\omega d}{2\pi - \Delta_d \phi}} = \frac{n(1+2\frac{h_1+2g+h_3}{d})}{\frac{f}{1-\Delta_d\phi/(2\pi)}}. \quad (B.1)$$

The formula given above yields the right time for such an instance, measuring from the start of rise of signal above noise, which occurs at 4.103 sec. The time duration from that instance to the proper instance of phase-difference check is calculated according to the formula, namely, as the acoustic distance divided by the acoustic speed. The acoustic distance is the travel distance of sound as it goes through the length of every currogation (top and bottom) on its way (n = 37 being the number of unit-cells from the speaker to the microphone after passing the entire waveguide and returnng from the open end back to the microphones – the best measurement time is just before the rerafaction wave returns to inform the microphone that the waveguide is truncated). The instance 4.102 sec is when the wave travelling directly through the solid body of the waveguide reaches the microphones directly and creates the first detectible signal: $dt = 15d/\sqrt{E/\rho} \sim 0.16$ msec << 7.5 msec, which means that this happens at some instance almost immedeately, setting the right moment to start counting the time till the signal through the acoustic distance in air reaches the microphones upon returning from the open end of the waveguide.

The time from the moment the signal reaches the microphone through the solid structure till the moment the rarefaction wave in air reaches the microphone depends not only on the acoustic distance but also on the acoustic speed. The acoustic (phase) speed at a given frequency of excitation depends on the phase. The pressure in the microphone matures for measurement when the slower of the two phases $\{\Delta_d\phi, 2\pi - \Delta_d\phi\}$ reaches the microphones. Since the phase difference of $\pi$ is never reached for relevant gas concentrations, the appropriate phase difference is $2\pi - \Delta_d\phi$. For the calculation of $\Delta t_\phi$ the phase difference is still unknown. In this sense it is an iterative process. First, the wavenumber should be assumed in relative-phase units. One can either just assume any number and repeat the process several times until convergence is reached. A shorter path is assuming the right wavenumber using the numerical dispersion diagram and the operational frequency.

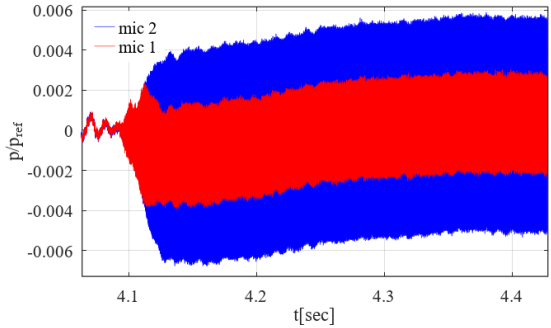

(a)

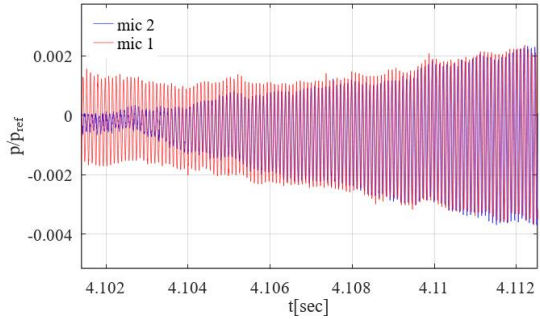

(b)

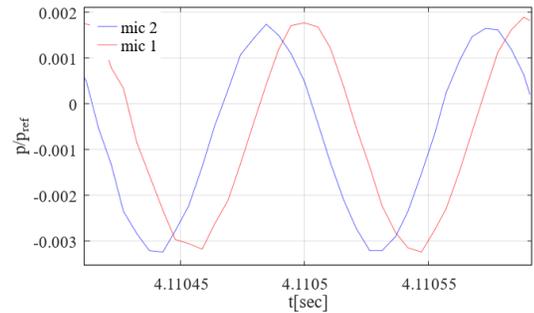

(c)

Figure B.1: Pressures histories for excitation frequency of 11.158kHz for concentration of 0.5% CO2 at different magnification resolutions (a), (b), and (c), showing the calculated phase $\Delta d\phi \sim \pi/3$ as shown in Fig. 10.

Using the theoretical diagram can only shorten the iterative process, by producing a better initial guess, but even without it the iterative process will converge. This is to say that the procedure is by no means one of cyclic logic.

Since the dependence of the phase difference on the right time instance for comparison is weak, when taking the initial guess for the wavenumber based on the computational plot, two iterations are sufficient.